\documentclass[a4paper,twocolumn,english,aps,prl,showpacs,floatfix]{revtex4}
\usepackage{graphicx}

\makeatletter

\usepackage{babel}
\makeatother
\begin{document}

\title{Quantum dwell times}

\author{J. A. Damborenea}

\affiliation{Fisika Teorikoaren Saila, Euskal Herriko Unibertsitatea, 644 P. K.,
48080 Bilbao, Spain}

\affiliation{Dpto. de Química Física, Universidad del País Vasco, Apdo. 644, 48080
Bilbao, Spain}

\author{I. L. Egusquiza}

\affiliation{Fisika Teorikoaren Saila, Euskal Herriko Unibertsitatea, 644 P. K.,
48080 Bilbao, Spain}

\author{J. G. Muga}

\affiliation{Dpto. de Química Física, Universidad del País Vasco, Apdo. 644, 48080
Bilbao, Spain}

\author{B. Navarro}

\affiliation{Dpto. de Química Física, Universidad del País Vasco, Apdo. 644, 48080
Bilbao, Spain}

\begin{abstract}
We put forward several inherently quantum characteristics of the dwell
time, and propose an operational method to detect them. The quantum
dwell time is pointed out to be a conserved quantity, totally bypassing
Pauli's theorem. Furthermore, the quantum dwell time
in a region for one dimensional motion is doubly degenerate. In presence
of a potential barrier, the dwell time becomes bounded, unlike the
classical quantity. By using off-resonance coupling to a laser we
propose an operational method to measure the absorption by a complex
potential, and thereby the average time spent by an incoming atom in the
laser region.
\end{abstract}

\date{10th March 2004}

\maketitle
Time has traditionally been a sore subject in quantum mechanics. Since
the early days of the theory, the inclusion of time observables in
the usual formalism has proved problematic (for a general presentation
of several aspects of time in quantum mechanics, see \cite{MSE02}).
The advent of single atom manipulation and fantastic cooling techniques,
however, requires further thinking on the quantum aspects of measurements
of time. In this letter we present an analysis of the dwell time observable,
in which we shall prove that some hitherto unexplored properties of
the self-adjoint operator are intrinsically quantum mechanical in
origin. We analyze the possiblity that these aspects could be experimentally
observed in two different setups, and realize that indeed current
techniques will allow us the exploration of quantum properties of
dwell time.

The concept of a {}``dwell time'' for a stationary regime was first
clearly made distinct from {}``traversal time'', {}``delay time''
and {}``reflection time'' by B\"uttiker \cite{Buttiker83}. Without
explicit mention of operators, similar quantities had been presented
in the literature previously as components of a {}``delay time'',
obtained by suitable substraction of the corresponding object for
the free case, and a limiting procedure, adequate for scattering processes
\cite{Wigner55,Smith60}. Much of the discussion of the dwell time
has concerned stationary states (see however \cite{BSM94,Nussenzveig00,JW87,JW89,HFF87,HS89,SB87,LA93,Leavens95}).
At any rate, the following (dwell time) operator was at least implicitly,
and often explicitely \cite{JW87}, part of those definitions:\begin{equation}
\hat{t}_{D}=\int_{-\infty}^{+\infty}d\tau\, e^{i\hat{H}\tau/\hbar}\chi_{D}(\hat{x})\, e^{-i\hat{H}\tau/\hbar}\,.\label{eq:definition}\end{equation}
In this expression, $\chi_{D}(\hat{x})$ is the projector on the region
of interest, in which we desire to understand and compute the time
a quantum particle spends. The main emphasis in the literature has
lied on the fact that the average value of this observable over any
given state involves contributions from both reflection and transmission
situations, as well as interference terms. Some striking facts about
this operator seem however to be missing from current discussions. 

In particular, the operator $\hat{t}_{D}$ is prima facie symmetric;
but it can further be proved that it is in fact essentially self-adjoint
in the cases we shall be considering. For free particles moving in
one dimension, the normalizability requirement of the image of the
operator determines the (initial) domain to be that of square summable
wave functions which, in momentum representation, fulfill $\psi(p)=o(\sqrt{p})$.
Symmetry imposes no further constraint on the domain, which is dense,
and the deficiency indices are computed to vanish. For the case of
a purely scattering spectrum, this proof is translated without further
ado. Additionally, notice that $\hat{t}_{D}$ is a positive definite
operator.

Over the common domains of definition, it is immediate to observe,
although not generally known, that the Hamiltonian commutes with the
dwell time operator. This should come as no surprise even to those
attuned to the general meaning of Pauli's theorem, which is concerned
with a \emph{covariant} time operator, i.e., an observation of an
\emph{instant}, since here we deal with an \emph{interval} observable.
This is in fact the reason for the Wigner-Smith delay time \cite{Wigner55,Smith60}
to be defined in terms of scattering data. As a consequence, it will
be of interest to diagonalize the dwell time operator in the eigenspaces
of the Hamiltonian. For one dimensional motion under a Hamiltonian
with no bound states, this leads generically to two different eigenvalues
for the dwell time operator in each energy eigenspace. Thus we are
faced with the following logical outcome: if we were to design a measuring
procedure for the dwell time observable that indeed coincides with
the dwell time operator presented above, the probability density for
measured dwell times would be generically bimodal for particles with
small energies (in fact, the requirement is that the characteristic
action, given by the characteristic momentum of the particle multiplied
by the length of the region in which we measure the permanence of
the particle, has to be much smaller than $\hbar$.) In an interval
of length $l$, the dwell time eigenvalues for a free particle in
one dimension with momentum $p$ are $t_{\pm}(p)=\frac{ml^{2}}{\hbar}\left(\frac{\hbar}{|p|l}\right)\left(1\pm\frac{\hbar}{pl}\sin\frac{pl}{\hbar}\right)$.
It should be observed that $t_{+}(p)$ grows without bound when $p\to0$. 

Furthermore, we can generically define the expected dwell time of
a stationary state for a particle in one spatial dimension as its
expected value in the two dimensional Hilbert space with the associated
energy; for instance, for a free particle of momentum $p$ the expected
value would be $\langle\tau_{D}\rangle=\frac{ml^{2}}{\hbar}\left(\frac{\hbar}{|p|l}\right)$,
which actually coincides with the classical quantity. 

As pointed out above, the eigenvalues of the dwell time operator are
not bound from above for the free particle case. That is not the situation
however when the particle faces a barrier. In this case both series
of eigenvalues are bounded, which entails that there is a maximum
dwell time for a quantum particle in a barrier, in sharp contrast
to the classical situation, where the dwell time for particles whose
energies allows them to overcome the barrier can be made as big as
desired by smoothly diminishing the energy. The preceding statements
can be proved as follows. First, observe that the matrix elements
of the projector of the region of interest in the scattering basis
$\left|p^{+}\right\rangle $ can be easily computed if it encompasses
completely the interaction region, since $\left\langle p^{+}\right|\chi_{D}(\hat{x})\left|q^{+}\right\rangle =\delta(p-q)-\left\langle p^{+}\right|\bar{\chi}_{D}(\hat{x})\left|q^{+}\right\rangle $,
where $\bar{\chi}_{D}(\hat{x})$ is the complementary projector, and
the explicit forms of the scattering eigenstates in the position representation
outside the interaction region are known in terms of the scattering
amplitudes. The possibly dangerous delta terms vanish in the end result
thanks to unitarity, which in fact underpins the whole construction,
and the explicit form of the eigenvalues of the dwell time operator
can be obtained in terms of the transmission and reflection amplitudes.
For the specific case of a square barrier of height $V_{0}$, the
eigenvalues are given by \[
t_{\pm}(p)=2ml|p|\frac{1\pm(\hbar/lq)\sin(lq/\hbar)}{p^{2}+q^{2}\pm2mV_{0}\cos(lq/\hbar)},\]
where now $q^{2}=p^{2}-2mV_{0}$. The maxima and minima of $t_{+}$ and
$t_{-}$ interleave, and both functions tend to the classical value
for high incident momenta. In the same way as above, one can define
the expected dwell time for a stationary state with momentum $p$,
which is given by $\langle\tau_{D}\rangle=(t_{+}(p)+t_{-}(p))/2$.
This average dwell time is in fact what has been usually termed {}``dwell
time'' in the literature \cite{Buttiker83,HS89}. The decomposition
of an average dwell time for a generic state in terms of a symmetric
and an antisymmetric part was suggested by Nussenzveig \cite{Nussenzveig00},
and it holds true for parity invariant hermitian Hamiltonians.

It now behooves us to propose an operational model to test the prediction
that dwell times will present, for a number of particle states, a
bimodal distribution. The first one that comes to mind is one based
on fluorescence, in analogy to a time of arrival detection through
fluorescence \cite{DEHM02,DEHM03}: consider that the region of interest
is illuminated by a laser on resonance with an internal transition
of the particle, perpendicular to the initial motion of the particle;
set this laser and the initial state so as to minimize motion transversal
to the classical path of the particle; then measure the number of
resonance photons emitted by the particle for each run of the experiment.
It could be expected that indeed the distribution of numbers of emitted
photons would be a proxy for the distribution of dwell times, and,
therefore, that for some regime this distribution of emitted photons
would also be bimodal. However, the regime of interest in which the
bimodal distribution would be observed is actually inadequate for
analysis with this operational model, since reflection and detection
delays will be present due to the laser \cite{DEHM02}, leading to
poor total signal and poor signal discrimination. In particular, for
the bimodality to be observed, the characteristic interval between
modes ($\hbar/E$, where $E$ is the particle's energy) should be
greater than the characteristic interval between successive emission
of fluorescence photons ($2/\gamma+\gamma/\Omega^{2}$, where $\gamma$
is the decay constant, i.e. Einstein's coefficient, and $\Omega$
Rabi's frequency); on the other hand, if reflection is to be avoided,
the particle energy should be much bigger than the characteristic
Rabi energy ($E\gg\hbar\Omega$). These conditions cannot be simultaneously
met, making this procedure inadequate for revealing the bimodality
of dwell times.

An operational method for determining the average dwell time does
indeed exist, using absorbing potentials \cite{GFGG90,HW91,MBS92a}.
Consider the non hermitian Hamiltonian $H_{{V_{I}}}$ obtained from
adding to a scattering Hamiltonian a complex term of the form $-i{V_{I}}\chi(\hat{x})$.
The evolution under this Hamiltonian will result in absorption; nonetheless,
it is possible to define transmission and reflection coefficients
in the usual manner. Following the method described by Smith \cite{Smith60}
to relate the delay time of a stationary state to the derivatives
of transmission and reflection coefficients, but in this case with
respect to ${V_{I}}$, we obtain the average dwell time in the region
of interest as $\langle\tau_{D}(p)\rangle=\lim_{{V_{I}}\to0}(\hbar/2)\partial_{{V_{I}}}A(p)$,
where $A(p)$ is the total absorption probability for incident momentum
$p$. The equivalence of this quantity with $(t_{+}(p)+t_{-}(p))/2$
can be readily checked. Since the absorption in a region can be tuned
in a number of manners, we can check the reality of the quantum prediction
at hand, differing from the classical one; namely, that for all ingoing
waves the quantum mechanical dwell time is bounded, unlike the classical
one.

\begin{figure}
\includegraphics[%
  width=1.0\columnwidth,
  keepaspectratio]{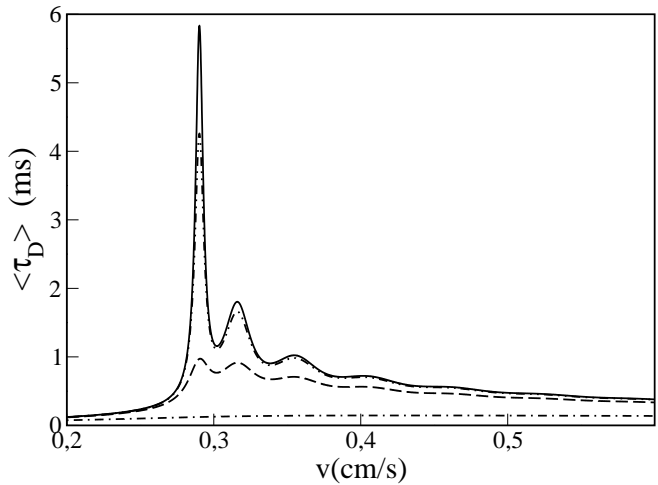}

\caption{\label{fig:figdwell6}Exact average dwell time (solid line) for Cs
atoms crossing a 2 $\mu$m barrier of height corresponding to 0.28
cm/s. Approximate values $\tau_{{\rm approx}}$ are given for (also
solid line) $\Delta=2500\gamma$, $\Omega=1.57\gamma$; (double dotted
- dashed line) $\Delta=250\gamma$, $\Omega=0.5\gamma$; (dashed line)
$\Delta=25\gamma$, $\Omega=0.16\gamma$; (dotted - dashed line) $\Delta=2.5\gamma$,
$\Omega=0.05\gamma$. In all cases the resonance is at the 852 nm
transition with $\gamma=33.3\times10^{-6}$s$^{-1}$. }
\end{figure}

In particular, consider a two level system coupled in a spatial
region to an off-resonance laser with large detuning, $\Delta\gg\gamma,\Omega$,
where $\Delta$ is defined as the laser frequency minus the frequency
of the atomic transition. The amplitude for the atomic ground state
up to the first photon detection is governed then by the effective
potential \cite{NEMH03a,MCH93} \[
V(x)=(V_{R}-iV_{I})\chi(x)=\left(\frac{\hbar\Omega^{2}}{4\Delta}-i\frac{\hbar\gamma\Omega^{2}}{8\Delta^{2}}\right)\chi(x),\]
so that the average detection delay is now $4\Delta^{2}/\Omega^{2}\gamma$.
Whereas $\gamma$ is fixed for the atomic transition, both $\Omega$
and $\Delta$ may be controlled experimentally. The ratio $\Omega^{2}/\Delta$
can always be chosen so that the real part of $V$ remains constant.
The remnant freedom can be used to set the value of the imaginary
part. We desire that at most one fluorescence photon is emitted per
atom, so that the fluorescence signal produced by an atomic ensemble
will be proportional to the absorption probability $A$. This requirement
can be met in the regime determined by $\langle\tau_{D}(p)\rangle\ll4\Delta^{2}/\Omega^{2}\gamma$;
that is to say when the average time spent in the region is much smaller
than the delay. After carrying out the adequate calibration to take
into account the detector solid angle and efficiency, succesive measurements
of $A$ are carried out for different (automatically small) values
of $V_{I}$. In this manner we can obtain an approximate value for
the derivative of the absorption probability with respect to the imaginary
part of the potential, and hence $\langle\tau_{D}(p)\rangle$. In
fact, it is easier to compute $\tau_{{\rm approx}}=\hbar A/(2V_{I})$,
for small values of $V_{I}$. The limitation inherent to the procedure
lies in the weak signal for small absorption, whereas strong absorption
leads to larger errors when using the approximate expression $\tau_{{\rm approx}}$.
However, for absorption of the order of 0.2 the relative error in
estimating $\langle\tau_{D}(p)\rangle$with $\tau_{{\rm approx}}$
is of the same order of magnitude. In fig. \ref{fig:figdwell6} we
depict the exact average dwell time as a function of incident velocity
for Cs atoms, as well as approximations given by $\tau_{{\rm approx}}$
for several values of $V_{I}$.

\begin{figure}
\includegraphics[%
  width=1.0\columnwidth]{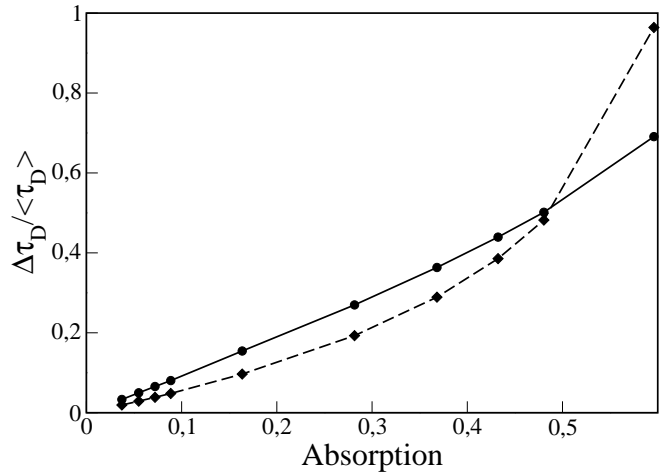}

\caption{\label{fig:figdwell11} Relative error (circles) and average dwell
time over fluorescence delay time (diamonds) as a function of absorption
for the same incidence conditions as in fig. \ref{fig:figdwell6}, at the largest peak thereof.}
\end{figure}

To assess better the weak signal limitation of the procedure, in fig.
\ref{fig:figdwell11} we depict both the relative error $(\langle\tau_{D}(p)\rangle-\tau_{{\rm approx}})/\langle\tau_{D}(p)\rangle$
and the quotient between the average dwell time and the fluorescence
delay time versus absorption, at the dwell time peak. The validity of
the approximation is controlled by the smallness of this latter
quotient.

In conclusion, we have signalled a number of hitherto unnoticed properties
of the dwell time operator, which are intrinsically quantum mechanical,
namely that it is a stationary observable and with double degeneracy
in one-dimensional collisions. We have analyzed two possible operational
approaches to unveil these properties, and we have detailed a procedure
making use of absorption that could experimentally lead to measurement
of atomic dwell times.

We acknowledge support by UPV-EHU (00039.310-13507/2001), Spanish
Ministry for Science and Technology and FEDER (BFM2003-01003 and
FPA2002-02037). JAD is supported by the Basque Department of
Education, Science and Research.
\bibliographystyle{apsrev}

\end{document}